\documentclass[twocolumn,preprintnumbers,amsmath,amssymb]{revtex4}
\usepackage{graphicx} % include figure files
\usepackage{adjustbox}
\usepackage{setspace} % include figure file
\usepackage{dcolumn} % align table columns on decimal point
\usepackage{bm} % bold math
\usepackage{natbib}
\usepackage{hyperref}
\hypersetup{colorlinks=true,linkcolor=blue,citecolor=blue,urlcolor=blue}
\usepackage{epsfig}

\begin{document}

\title{Magnetic and Electrical Properties of high-entropy rare-earth manganites}

\author{Ashutosh Kumar$^{a}$}
\author{David Bérardan$^a$}
\author{Diana Dragoe$^a$}
\author{Eric Riviere$^a$}
\author{Tomohiro Takayama$^{b,c}$}
\author{Hidenori Takagi$^{b,c}$}
\author{Nita Dragoe$^{a,}$\footnote{Email: nita.dragoe@universite-paris-saclay.fr}}

\affiliation{$^a$ICMMO (UMR CNRS 8182), Université Paris-Saclay, F-91405 Orsay, France}

\affiliation{$^b$Max Planck Institute for Solid State Research, Heisenbergstrasse 1, 70569 Stuttgart, Germany}

\affiliation{$^c$Institute for Functional Matter and Quantum Technologies, University of Stuttgart, Pfaffenwaldring 57, 70550 Stuttgart, Germany}
\date{\today}
\begin{abstract}
 Detailed investigations of structural, magnetic and electronic transport properties of hole-doped high-entropy rare-earth manganites are presented. The high-entropy samples (LaNdPrSmEu)$_{1-x}$Sr$_x$MnO$_3$ (0$\leq$\textit{x}$\leq$0.5), synthesized using the solid-state technique, show a change in the crystal structure from \textit{Pbnm} to \textit{R-3c} with increasing Sr substitution, attributed to the change in the tolerance factor. Prominent ferromagnetic ordering is observed in the sample with a rhombohedral structure (\textit{x}$\geq$0.3), originating from the dominant double exchange mechanism mediated by itinerant electrons. Further, the Curie temperature is smaller for the high-entropy sample with \textit{x}=0.3, as compared to La$_{0.7}$Sr$_{0.3}$MnO$_3$, suggesting a strong relation between the Curie temperature and the Mn-O-Mn bond angle associated with the reduced ionic radii at the rare-earth site. The electrical resistivity of the high-entropy samples is larger than those of La$_{1-x}$Sr$_x$MnO$_3$, which can be ascribed to the reduced bandwidth due to the enhanced structural distortion. A concomitant rise in magnetoresistance is observed for high-entropy samples with the increase in Sr concentration. These findings considering the configurational complexity of different rare-earths advance the understanding of high-entropy rare earth manganites.\\    
\end{abstract}
\maketitle
\section{Introduction}
High-entropy oxides (HEOx) are emerging as an advanced research area for multifunctional applications since their first discovery with (MgCoNiCuZn)O rock-salt oxide.\cite{R1} The possibility of adding several elements randomly at a single crystallographic site enhances the configurational entropy (maximum when five or more elements are added in equimolar proportion).  The increased configurational entropy, in some cases, results in the structural stabilization of a compound which is referred to as the entropy-stabilized phase.\cite{R2} The increased configurational entropy has proven to induce interesting exotic physical properties including long-range antiferromagnetism\cite{R3, R4}, thermoelectrics\cite{R5, R6}, ionic conductivity\cite{R7, R8}, colossal dielectric constant\cite{R9}, proton conductivity\cite{R10}, etc. More generally HEOx, even with non-entropy-stabilized phases, constitute an extraordinary playground to study the influence of local structural and chemical disorder on the functional properties. These exciting properties in HEOx have strongly invigorated the scientific community to understand their fundamental behavior.\cite{R11, R12}\\ 
Magnetic and electric transport properties of simple perovskite oxides (having general formula ABO$_3$; A=RE elements, B= transition metals, O is oxygen) are, in general, strongly related to the spin and orbital configurations and have been thoroughly studied in literature.\cite{R13} Among these, rare-earth (RE) manganites are an exciting class of materials with promising physical properties including colossal magnetoresistance (CMR), ferromagnetism, metallicity, metal-to-insulator transition (MIT), charge-, spin-, and orbital-ordering phenomena, etc.\cite{R14, R15, R16, R17, R18} These properties are influenced by the lattice distortion, Jahn-Teller distortion of Mn$^{3+}$ ions, and formation of Mn$^{4+}$ via divalent cations (Sr, Ba, Ca, etc) substitutions at the rare-earth site.\cite{R13} In particular, the Curie temperature (T$_C$) in the rare-earth manganites depends on the Mn-O-Mn bond angle\cite{R13}, which is characterized by the Goldschmidt tolerance factor ($t$). The tolerance factor  $t$=$\frac{(r_A+r_O)}{\sqrt{2}(r_B+r_O)}$, in perovskite depends on the average ionic radii of the A-site (r$_A$), B-site (r$_B$), and oxygen (r$_O$). The crystal structure stability is largely associated with $t$, which varies with the cationic radii. Also, the crystal structure in these rare-earth transition metal perovskite oxide materials has a significant influence on their unique physical properties.\cite{R19} The pristine single-A-cation rare-earth manganites (La/Nd/Pr/Sm/Eu)MnO$_3$ are antiferromagnetic insulators at low temperatures. However, they exhibit different magnetic transition temperatures depending on their ionic radii that change the Mn-O-Mn bond angle, which results in a significant change in hybridization between Mn-3d and O-2p orbitals states.\cite{R13}\\ 
High-entropy oxide materials are, in general, designed using an equimolar ratio of cations at the A-site or B-site, and several intriguing magnetic properties have been investigated in recent times. For example, an enhanced magnetic frustration has been demonstrated in (MgCoNiCuZn)Al$_2$O$_4$ [20], and a spin-glass ground state has been observed in high-entropy rare-earth pyrochlore (Yb$_{0.2}$Tb$_{0.2}$Gd$_{0.2}$Dy$_{0.2}$Er$_{0.2}$)$_2$Ti$_2$O$_7$.\cite{R21} Weak ferromagnetism at T$<$10\,K has been observed in high-entropy rare-earth cobaltate (Gd$_{0.2}$Nd$_{0.2}$La$_{0.2}$Sm$_{0.2}$Y$_{0.2}$)CoO$_3$.\cite{R22} Single-phase high-entropy perovskites with an equimolar amount of elements at both the A and B sites have been synthesized.\cite{R23, R24} Witte et al. reported the detailed magnetic properties of high-entropy rare-earth transition metal perovskite oxides that show predominant antiferromagnetic behavior with minor ferromagnetic contributions in pristine samples.\cite{R25} The impact of different A-site cations on the electrical properties of high-entropy LnCr$_{0.2}$Mn$_{0.2}$Fe$_{0.2}$Co$_{0.2}$Ni$_{0.2}$O$_3$ has been described recently.\cite{R26} The substitution at the RE site with divalent alkaline elements has shown promise in the past to drastically alter the electronic and magnetic properties in simple rare-earth perovskites. However, it has been barely studied in high-entropy perovskites. In this work, the effect of Sr-substitution at the RE site on the electronic and magnetic properties in high-entropy rare-earth manganite (LaNdPrSmEu)$_{1-x}$Sr$_x$MnO$_3$ (0$\leq$\textit{x}$\leq$0.5) is discussed, and a comparison with simple perovskite is presented.
\section{Experimental Work}
High-entropy rare-earth manganites (LaNdPrSmEu)$_{1-x}$Sr$_x$MnO$_3$ (0 $\leq$ \textit{x} $\leq$ 0.5) were synthesized using standard solid-state reaction methods with an equimolar ratio of the rare-earth elements. Stoichiometric amounts of La$_2$O$_3$ (Alfa Aesar, 99.9\%), Nd$_2$O$_3$ (Alfa Aesar, 99.9\%), Pr$_2$O$_3$ (Alfa Aesar, 99.9\%), Sm$_2$O$_3$ (Alfa Aesar, 99.9\%), Eu$_2$O$_3$ (Alfa Aesar, 99.9\%), SrCO$_3$ (Alfa Aesar, 99.9\%) and MnO$_2$ (Alfa Aesar, 99.99\%) were mixed using planetary ball milling (Fritsch Pulverisette 7 Premium Line) at 350 rpm for 12 cycles (5 min on and 1 min off time). The rare-earth precursors were heated overnight at 1173\,K before synthesis to avoid any water adsorption in the powders. The mixed precursors were heated at 1573\,K for 24 hours in alumina crucibles with a heating and cooling rate of 200 K/hour. The calcined powders were further consolidated into bar-shaped pellets and sintered at 1623\,K for 20 hours in alumina crucibles with a 200\,K/hour heating, and cooling rate. The structural characterization of the samples was performed by X-ray diffraction using a Panalytical X'Pert diffractometer with a Ge(111) incident monochromator, a copper tube (K$_{\alpha1}$ radiation), and a fast detector (X'celerator). Further, X-ray photoelectron spectroscopy (XPS) measurements were performed on a Thermo Fisher Scientific instrument with a monochromatic Al-K$_\alpha$ X-ray source (energy 1486.68\,eV) and a hemispherical analyzer. Samples were measured in powder form. The base pressure was around 5$\times$10$^{-9}$ mbar and the diameter of the X-ray beam spot was 400\,$\mu$m, corresponding to an irradiated surface of approximately 1 mm$^2$. The hemispherical analyzer was operated at 0$^{\circ}$ take-off angle in the Constant Analyzer Energy (CAE) mode. Wide scan spectra were recorded at pass energy of 200\,eV and an energy step of 1\,eV while narrow scan spectra were recorded at pass energy of 50\,eV and an energy step of 0.1\,eV. Charge compensation was achieved employing a “dual beam” flood gun, using low-energy electrons ($<$5 eV) and argon ions. The binding energy scale was calibrated on the neutral carbon set at 285\,eV. The obtained core-level spectra were treated using CasaXPS software.\cite{R27} The electrical resistivity was measured using the standard four-probe technique using a physical properties measurement system (PPMS, Quantum Design) over a wide temperature range from 2\,K to 300\,K under 0\,T and 6\,T magnetic fields. The magnetic measurements (Magnetizaton vs Temperature (M-T) curves in zero field cooling (ZFC) and field cooling (FC) processes at 100\,Oe, and magnetization vs magnetic field (M-H) curve at 2\,K under FC (100 Oe) with a field sweeping of $\pm$5\,T) were performed using a SQUID magnetometer, (MPMS, Quantum Design, USA).\\
\section{Results and Discussion}
\begin{figure}
\centering
\includegraphics[width=0.98\linewidth]{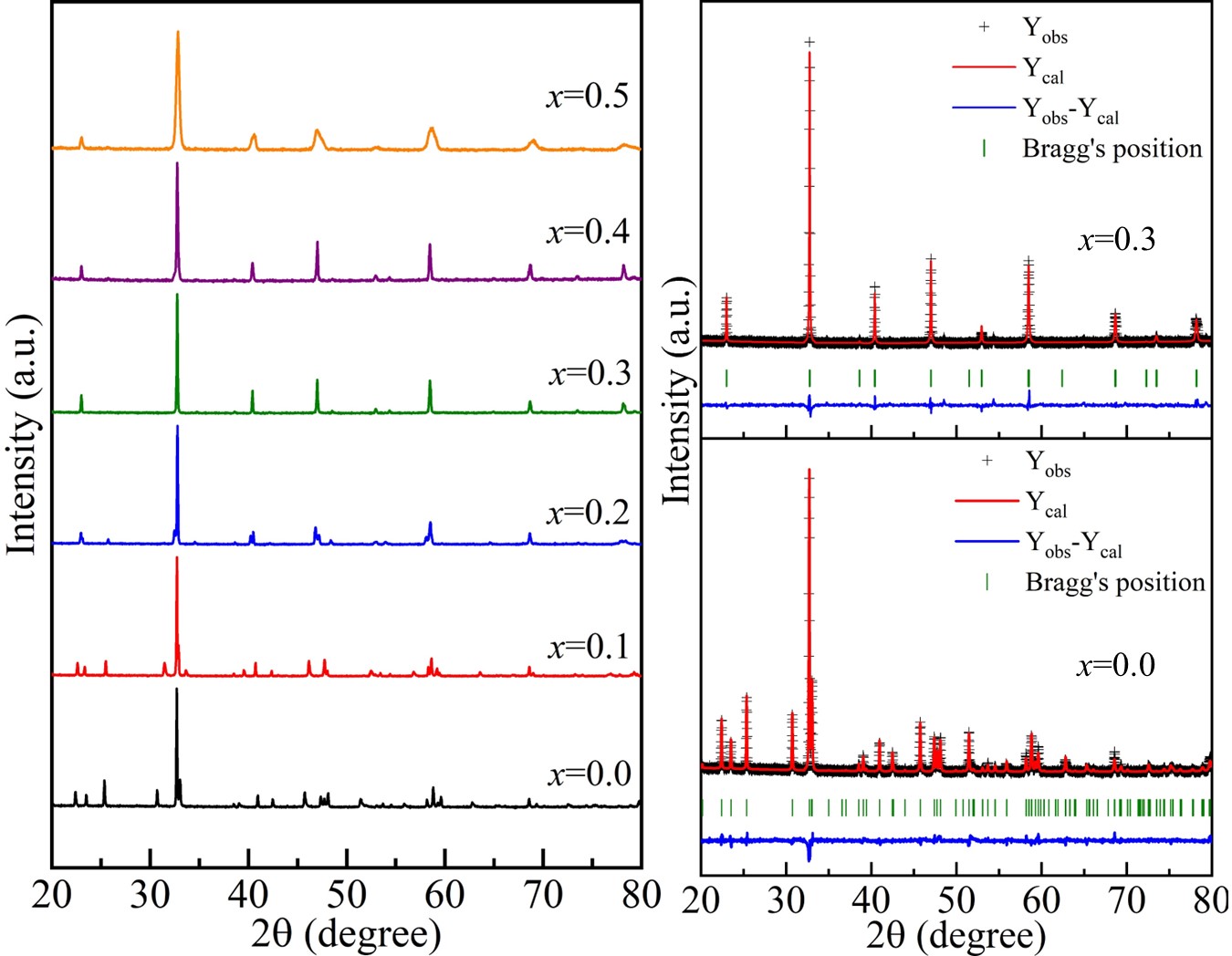}
\caption{XRD pattern for (LaNdPrSmEu)$_{1-x}$Sr$_x$Mn$O_3$ (0 $\leq$ \textit{x} $\leq$ 0.5). The Rietveld refinement pattern for \textit{x}=0.0 (Orthorhombic, \textit{Pbnm}) and \textit{x}=0.3 (Rhombohedral, \textit{R-3c}) is shown.}
\end{figure}
\begin{figure*}
\centering
\includegraphics[width=0.95\linewidth]{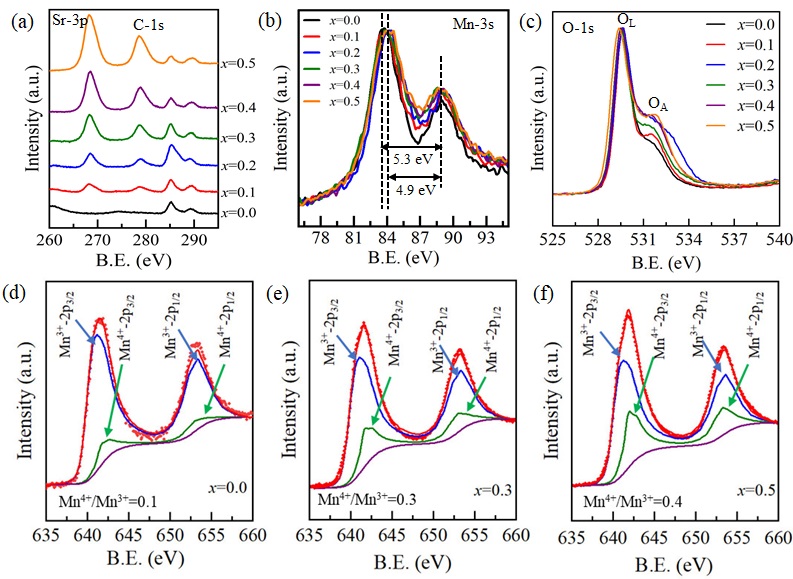}
\caption{Core-level X-ray photoelectron spectroscopy (XPS) spectra for  (LaNdPrSmEu)$_{1-x}$Sr$_x$MnO$_3$ (0$\leq$\textit{x}$\leq$0.5) samples (a) Sr-3p, (b) Mn-3s (c) O-1s and Mn-2p for (d) \textit{x}=0.0. (e) \textit{x}=0.3, and (f) \textit{x}=0.5.}
\end{figure*}
\subsection{Structural Properties}
The powder X-ray diffraction (XRD) patterns of the high-entropy (LaNdPrSmEu)$_{1-x}$Sr$_x$MnO$_3$ (0$\leq$\textit{x}$\leq$0.5) samples are shown in Fig.~1. The samples with \textit{x}=0.0, 0.1, and 0.2 can be successfully indexed with an orthorhombic structure (\textit{Pbnm}, Z=4)\cite{R28}, which clearly indicates a single-phase formation of these samples without any sign of impurity phases within the sensitivity of the XRD. Further Sr substitutions (\textit{x}=0.3, 0.4, and 0.5) lead to a rhombohedral distorted perovskite structure (\textit{R-3c}, Z=2). The change of crystal structure from orthorhombic to rhombohedral with Sr substitution is attributed to the ionic radii of Sr ions, as observed in simple rare-earth manganites.\cite{R29, R30} In the present study, the rare-earth site (A-site) in the ABO$_3$ structure is occupied by five different rare-earth elements (La, Nd, Pr, Sm, Eu) and Sr as a divalent alkaline element. The Goldschmidt tolerance factor ($t$) for the pristine sample (\textit{x}=0.0) is 0.92 and it changes to 0.95 for \textit{x}=0.5 using Shannon’s ionic radii.\cite{R31} This change in $t$ explains the structural change from orthorhombic to rhombohedral.\\
The powder XRD patterns have been further analyzed using Rietveld refinement employing Fullprof software.\cite{R32} The refinements have been performed with different space groups, depending on the Sr concentration in the system. The structural parameters obtained from the refinement are shown in Table~SI. The corresponding refinement patterns for (LaNdPrSmEu)$_{1-x}$Sr$_x$MnO$_3$ with \textit{x}=0.0 and 0.3 are shown in Fig.~1. The refinements for these samples converged well and gave decent goodness of fit, confirming that the samples are single-phase and well crystallized. However, it is noted that (LaNdPrSmEu)$_{1-x}$Sr$_x$MnO$_3$ with \textit{x}=0.5 does not fit accurately with the \textit{R-3c} structure since every peak has a broad tail (shown in Fig.~S1). This broadening in the XRD pattern could constitute the signature of phase separation in half-doped manganites.\cite{R33} To summarize, the change in crystal structure from \textit{Pbnm} to \textit{R-3c} originating from Sr substitution changes the Mn-O octahedra environment and hence may influence the physical properties including magnetism and resistivity.\cite{R34}\\

\subsection{X-ray photoelectron spectroscopy (XPS)}
\begin{figure*}
\centering
\includegraphics[width=0.95\linewidth]{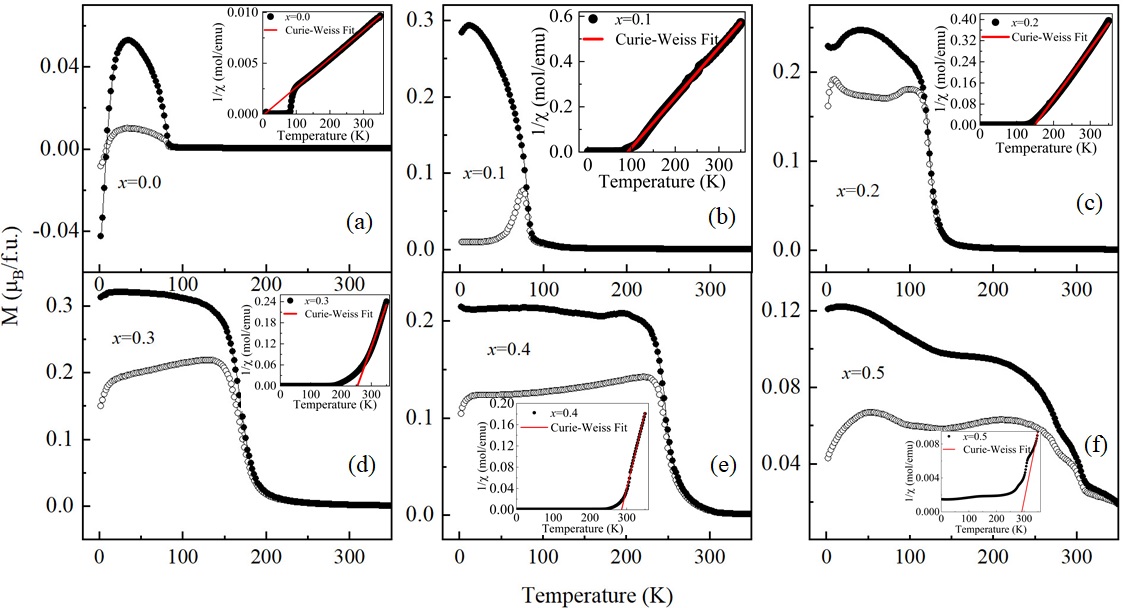}
\caption{Magnetization vs. temperature under ZFC (open symbol) and FC (closed symbol) mode under 100\,Oe for (LaNdPrSmEu)$_{1-x}$Sr$_x$MnO$_3$ (0 $\leq$ \textit{x} $\leq$ 0.5). The corresponding 1/$\chi$ vs temperature (T) is shown in the inset of each sample. The solid line presents the Curie-Weiss law fit.}
\end{figure*}
X-ray photoelectron spectroscopy (XPS) studies have been performed to investigate the change in the valence states of the elements present in the high-entropy (LaNdPrSmEu)$_{1-x}$Sr$_x$MnO$_3$ (0$\leq$\textit{x}$\leq$0.5) samples with the increase in Sr concentration, as shown in Fig.~2. The peak intensities were normalized with respect to the Mn-2p line intensity. The presence of all the expected elements is seen in the full-scan range of XPS (Fig.~S2). Fig.~2(a) shows the core-level spectrum for the Sr-3p peak, indicating the increased presence of Sr, which is consistent with the rise in Sr concentration in the samples. The influence of Sr substitution at the rare-earth site on the Mn valence state is seen in Fig.~2(b) and Fig.~2(d-f). The separation energy between the doublet peak of Mn-3s\cite{R35} indicates the rise in the Mn$^{4+}$ charge state in the system with Sr substitution (Fig.~2(b)). Further, the core-level spectra for Mn-2p indicate the presence of Mn-2p$_{3/2}$ at 643.6 $\pm$ 0.1 eV and Mn-2p$_{1/2}$ at 652.8 $\pm$ 0.1 eV due to spin-orbit coupling (Fig.~S3). Fig.~2(c) presents the core-level O-1s spectrum for different Sr concentrations. The O-1s intensity is fitted with oxygen contributions from the lattice site (O$_L$) and the environment with oxygen vacancies (O$_A$). It is seen that the O$_L$ contribution remains the same for all the samples; however, O$_A$ increases with increasing Sr concentration and may be attributed to a possible evolution of the concentration of oxygen vacancies due to the charge compensation effect.\\ 
The individual contributions of Mn$^{3+}$ and Mn$^{4+}$ are estimated from the core-level spectra of Mn-2p by fitting lineshapes derived from reference spectra (pure, well-characterized MnO$_2$ and Mn$_2$O$_3$ polycrystalline samples) according to a method described in ref \cite{R36} (Fig.~2(d-f)). A Shirley background was subtracted from all Mn-2p core lines. This approach simplifies the fitting routine while preserving real information, in contrast to the classical fitting routine that employs multiple synthetic line-shapes and many constraints, which can sometimes be arbitrary, especially in the case of transition elements.\cite{R37} It is seen that the Mn$^{4+}$/Mn$^{3+}$ ratio increases with increasing Sr concentration, which indicates that Sr substitution at the rare-earth site results in the formation of Mn$^{4+}$ in the system. Besides, it is noteworthy that the sample \textit{x}=0.0 already contains a small amount of Mn$^{4+}$, despite the absence of Sr.\\ 
\subsection{Magnetic Properties}
\begin{figure*}
\centering
\includegraphics[width=0.95\linewidth]{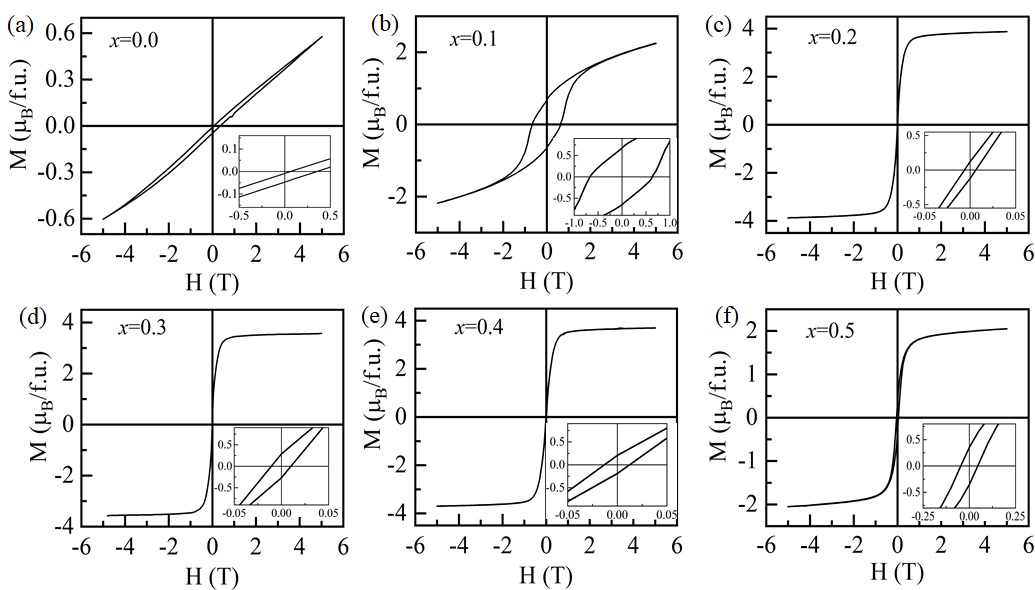}
\caption{Magnetization (M) vs. magnetic field (H) at 2\,K for (LaNdPrSmEu)$_{1-x}$Sr$_x$MnO$_3$. The inset shows the corresponding zoom-in M-H curve.}
\end{figure*}

The magnetization measurements as a function of temperature (M-T) at 100\,Oe magnetic field under field cooled (FC) and zero-field cooled (ZFC) processes for (LaNdPrSmEu)$_{1-x}$Sr$_x$MnO$_3$ (0$\leq$\textit{x}$\leq$0.5) are shown in Fig.~3. For all samples, the magnetization measurements indicate a magnetic transition upon cooling from 350\,K to 2\,K. It is noted that the individual rare-earth manganites display antiferromagnetic (AFM) ordering with the transition temperature varying from $\sim$ 145\,K for LaMnO$_3$ to 50\,K for EuMnO$_3$, attributed to the decrease in ionic radii that results in decreased Mn-O-Mn bond angle. However, as it can be observed in Fig.~3(a), the pristine (LaNdPrSmEu)MnO$_3$ sample displays a very different behavior. Indeed, a sharp increase of magnetization below 80\,K rather constitutes a possible signature of a ferro/ferrimagnetic transition with the Curie temperature (T$_C$) $\sim$ 80\,K. Below T$_C$, unexpected magnetic behavior is observed that may be related to the antiferromagnetic coupling of Mn ions to RE magnetic moments. Also, the random distribution of rare-earth elements may play a significant role in such exotic properties. Thus, (LaNdPrSmEu)MnO$_3$ may possibly be described as ferrimagnetic, where the ferromagnetic (FM) Mn lattice is oriented in the opposite direction to the average magnetic moment of the rare-earth elements (La, Nd, Pr, Sm, Eu). The FM ordering of Mn lattice for (LaNdPrSmEu)MnO$_3$ may be either ascribed to the canting of Mn$^{3+}$ moment due to Dzyaloshinsky-Moriya interaction as observed in
LaMnO$_3$ \cite{R37a, R37b} or probably to the double-exchange (DE) mechanism between Mn$^{3+}$-O$^{2-}$-Mn$^{4+}$ due to the presence of Mn$^{4+}$, as seen from the XPS analysis (for \textit{x}=0). Also, the superexchange (SE) mechanism between isovalent magnetic ions (Mn$^{3+}$-O$^{2-}$-Mn$^{3+}$ and Mn$^{4+}$-O$^{2-}$-Mn$^{4+}$) may contribute and hence competition between DE and SE interaction could result in a small magnetic moment. Upon further cooling, the magnetization is decreased and eventually shows negative values both in the ZFC and FC data at lower temperatures ($<$10\,K) for (LaNdPrSmEu)MnO$_3$. It could be due to decreased ionic radii for (LaNdPrSmEu)MnO$_3$ compared to LaMnO$_3$ which results in more buckling of the Mn-O-Mn bond angle hence reducing the strength of magnetic interaction. Thus, the ferromagnetic moment of the Mn lattice could be compensated by the paramagnetic nature of rare-earth elements used in high-entropy configuration aligning opposite to the applied magnetic field direction, and hence a magnetization reversal is observed. Similar reversal of magnetization is observed in RMnO$_3$ with R=Sm \cite{SmMnO3}, R=Gd \cite{GdMnO3}, R=Nd \cite{R40} and ascribed to the Mn-spin reorientation. Additionally, magnetization reversal has been observed in the manganite nanoparticles and discussed to originate from the surface effect.\cite{negativemagnetization} Such surface effects might be involved in the high-entropy sample. Further with an increase in Sr concentration at the A-site, the average magnetic moments of the rare-earth elements decrease and result in non-negative magnetization at the lower temperatures.\\
\begin{table*}[ht]
    \centering
    \caption{A comparison of magnetic transitions obtained in the individual rare-earth manganites with the present high-entropy (LaNdPrSmEu)$_{1-x}$Sr$_x$MnO$_3$ sample. (Neel temperature: T$_N$, Curie Temperature: T$_C$, Spin Glass Temperature: T$_{sg}$).}
  \begin{adjustbox}{width=1\textwidth}
    \small
      \begin{tabular}{c|c|c|c|c|c|c}
      Sample & La$_{1-x}$Sr$_x$MnO$_3$ & Nd$_{1-x}$Sr$_x$MnO$_3$ & Pr$_{1-x}$Sr$_x$MnO$_3$ & Sm$_{1-x}$Sr$_x$MnO$_3$ & Eu$_{1-x}$Sr$_x$MnO$_3$ & (LaNdPrSmEu)$_{1-x}$Sr$_x$MnO$_3$\\
       Ref.  & \cite{R13} & \cite{R13} & \cite{R13} & \cite{R46} &\cite{R47, R48} & \textbf{This work}\\
        \textit{x}=0.0  & T$_N$$\sim$145\,K & T$_N$$\sim$120\,K &	T$_N$$\sim$100\,K & T$_N\sim$60\,K &	T$_N$$\sim$50\,K &	T$_C$$\sim$80\,K\\
         \textit{x}=0.1  & T$_N$$\sim$130\,K & T$_N$$\sim$95\,K &	T$_C$$\sim$130\,K & T$_N$$\sim$60\,K &	-- &	T$_C$$\sim$82\,K\\
           \textit{x}=0.2 & T$_N$$\sim$280\,K & T$_C$$\sim$150\,K &	T$_C$$\sim$150\,K & T$_N$$\sim$60\,K &	T$_C$$\sim$65\,K &	T$_C$$\sim$122\,K\\
            \textit{x}=0.3 & T$_C$$\sim$370\,K & T$_C$$\sim$210\,K &	T$_C$$\sim$210\,K & T$_N$$\sim$60\,K &	T$_C$$\sim$62\,K &	T$_C$$\sim$174\,K\\
             \textit{x}=0.4 & T$_C$$\sim$370\,K & T$_C$$\sim$280\,K &	T$_C$$\sim$280\,K & T$_N$$\sim$60\,K &	T$_C$$\sim$65\,K &	T$_C$$\sim$245\,K\\
              \textit{x}=0.5 & T$_C$$\sim$360\,K & T$_C$$\sim$260\,K &	T$_C$$\sim$260\,K & T$_N$$\sim$60\,K &	T$_{sg}$$\sim$50\,K &	Multiple transitions\\
      \end{tabular}
   \end{adjustbox} 
    \label{tab:my_label}
\end{table*}
By substituting Sr ions at the A-site, the magnetic ground state changes from a ferrimagnetic to a ferromagnetic state, indicating the dominant appearance of double-exchange interactions mediated by itinerant Mn e$_g$ electron. Although the magnetization reversal disappears by Sr doping, significant bifurcation is seen between the ZFC and FC data in all the samples, which is likely associated with the freezing of (weak) ferromagnetic domains. The (LaNdPrSmEu)$_{0.9}$Sr$_{0.1}$MnO$_3$, as shown in Fig.~3(b), displays a much larger magnetization compared to (LaNdPrSmEu)MnO$_3$, suggesting the existence of ferromagnetic interaction, while the ordering temperature is similar to it. As described later (Fig.~4(b)), this sample shows a signature of competition between antiferromagnetic and ferromagnetic interactions, where the former is derived from the super-exchange (SE) process between isovalent Mn ions, rather than predominant ferromagnetic interactions. Upon further increasing the Sr content, the magnetic transition temperature is strongly enhanced, implying the change of dominant interaction. The (LaNdPrSmEu)$_{0.8}$Sr$_{0.2}$MnO$_3$ shows a magnetic transition at T$_C$ $\sim$122\,K (Fig.~3(c)). Further, the samples with higher Sr concentrations ((LaNdPrSmEu)$_{1-x}$Sr$_x$MnO$_3$ with \textit{x}=0.3, 0.4) show dominant ferromagnetic interactions and the magnetization of the samples with higher Sr-concentration (\textit{x}=0.3 (Fig.~3(d)), 0.4 (Fig.~3(e))) steeply increases at T$_C$ and remain almost constant at lower temperatures, suggesting a bulk ferromagnetic state. This change of magnetism is accompanied by the structural change from the orthorhombic to rhombohedral structure, which is consistent with the report in literature for single A-site rare-earth manganite.\cite{R42} (LaNdPrSmEu)$_{0.5}$Sr$_{0.5}$MnO$_3$ sample shows multiple magnetic transitions (Fig.~3(f)) and may be attributed to the phase separation phenomenon in half-doped manganites.\cite{R33} At low temperatures below 10\,K, weak anomalies are seen in the M-T curves, which could be ascribed to the AFM ordering among the rare-earth spins. The decrease of FC magnetization could indicate that the rare-earth elements and Mn sub-lattices interact antiferromagnetically.\\
The magnetic transition temperature (Curie temperature: T$_C$) obtained from the inflection point of the M-T curve (ZFC) for (LaNdPrSmEu)$_{1-x}$Sr$_x$MnO$_3$ (0$\leq$\textit{x}$\leq$0.5) is shown in Table~I. Also, it is noted that the magnetic transition is sharp in all the samples (except for (LaNdPrSmEu)$_{1-x}$Sr$_x$MnO$_3$ with \textit{x}=0.5) indicating the homogenous composition of the sample.\cite{R43} Magnetic transition temperatures observed in the high-entropy samples are lower than that of single-A-site rare-earth (La, Nd, and Pr) manganite for the same Sr substitution. The presence of rare-earth elements with smaller ionic radii results in enhanced tilting and rotation distortion of MnO$_6$ octahedra which reduces the Mn-O-Mn bond angle and a decrease in the spatial overlap of the Mn-3d orbital and O-2p orbital. This reduces the electron bandwidth which further attenuates the strength of the double exchange-ferromagnetic (DE-FM) interaction between the Mn$^{3+}$ and Mn$^{4+}$ ions thereby causing a lowering in magnetic transition temperatures. Therefore, the lower magnetic transition temperature obtained in the present study compared to LaMnO$_3$ is ascribed to the reduced ionic radii of the A-site than La as suggested elsewhere.\cite{R44, R45} The temperature dependences of the inverse susceptibility 1/$\chi$ for all the samples are shown in their respective inset and are fitted using the Curie-Weiss law: 1/$\chi$=(T-$\theta_p$)/C. The 1/$\chi$ fit shows a significant deviation above the transition temperatures for (LaNdPrSmEu)$_{0.7}$Sr$_{0.3}$MnO$_3$ (240\,K, T$_C$=174\,K) and (LaNdPrSmEu)$_{0.6}$Sr$_{0.4}$MnO$_3$ (280\,K, T$_C$=245\,K). It can be attributed to the development of short-range magnetic correlations, where long-range ordering may be partly prevented due to the inhomogeneities associated with the high-entropy configuration.\\ 
The magnetization vs magnetic field (M-H) curves for (LaNdPrSmEu)$_{1-x}$Sr$_x$MnO$_3$ (0$\leq$\textit{x}$\leq$0.5) at 2\,K are shown in Fig.~4(a-f). The sample was cooled under a small magnetic field of 100 Oe from 350\,K to 2\,K and then a sweeping magnetic field of $\pm$5T is used. The M-H curve for the samples with \textit{x}=0.0 (Fig.~4(a))and 0.1 (Fig.~4(b)) at 2\,K does not show saturation magnetization even at $\pm$5T and is indicative of the canted antiferromagnetic properties, as seen from the M-T curve. The inset shows the zoom-in M-H curve for (LaNdPrSmEu)MnO$_3$, suggesting an obvious field offset in both vertical and horizontal directions. Further with an increase in Sr concentration the magnetization in the system increases. For (LaNdPrSmEu)$_{1-x}$Sr$_x$MnO$_3$ with \textit{x}=0.1, the magnetization reaches a maximum value of 2.1 $\mu_B$/f.u, however, it does not reach saturation even at $\pm$5T. Also, a large hysteresis present in the M-H curve (shown in the inset of Fig.~4(b)) may be due to the competing antiferromagnetic (AFM) and FM interaction with the increase in Mn$^{4+}$ content with Sr substitution. Further rise in Sr concentration results in saturation magnetization for (LaNdPrSmEu)$_{0.8}$Sr$_{0.2}$MnO$_3$ (3.7 $\mu_B$/f.u) (Fig.~4(c)), (LaNdPrSmEu)$_{0.7}$Sr$_{0.3}$MnO$_3$ (3.2 $\mu_B$/f.u.) (Fig.~4(d)), and (LaNdPrSmEu)$_{0.6}$Sr$_{0.4}$MnO$_3$ (3.8 $\mu_B$/f.u.) Fig.~4(f). It further reduces to 2.1 $\mu_B$/f.u. for (LaNdPrSmEu)$_{0.5}$Sr$_{0.5}$MnO$_3$ (Fig.~4(f)). The precipitous reduction of the saturation magnetization for (LaNdPrSmEu)$_{0.5}$Sr$_{0.5}$MnO$_3$ is a signature of electronic and magnetic inhomogeneity induced by chemical disorder, as suggested by the broad tails in the XRD patterns. The zoom-in M-H curve for all the samples is shown in their corresponding insets, which shows a symmetric soft FM nature for the samples with higher Sr concentrations.\\

\subsection{Electrical Properties}
\begin{figure}
\centering
\includegraphics[width=0.9\linewidth]{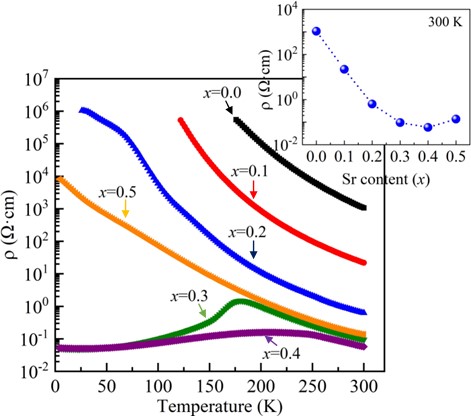}
\caption{Electrical resistivity ($\rho$) as a function of temperature for (LaNdPrSmEu)$_{1-x}$Sr$_x$MnO$_3$ (0 $\leq$ \textit{x} $\leq$ 0.5). Inset shows the dependence of $\rho$ vs Sr concentration in (LaNdPrSmEu)$_{1-x}$Sr$_x$MnO$_3$ (0 $\leq$ \textit{x} $\leq$ 0.5) at 300\,K}
\end{figure}
\begin{figure*}
\centering
\includegraphics[width=0.95\linewidth]{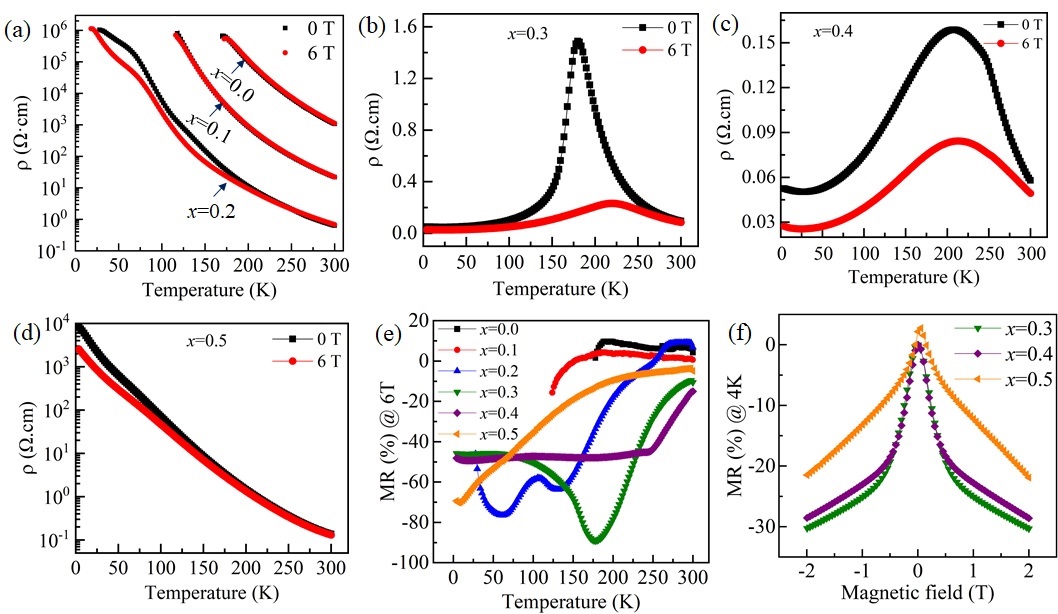}
\caption{Electrical resistivity ($\rho$) as a function of temperature (T) under 0\,T and 6\,T for (LaNdPrSmEu)$_{1-x}$Sr$_x$MnO$_3$ (a) \textit{x}=0.0, 0.1, 0.2 (b) \textit{x}=0.3, (c) \textit{x}=0.4, (d) \textit{x}=0.5. Magnetoresistance (MR(\%)) as a function of temperature, and (f) MR(\%) at 4\,K vs magnetic field for (LaNdPrSmEu)$_{1-x}$Sr$_x$MnO$_3$ with \textit{x}=0.3, 0.4 and 0.5.}
\end{figure*}
Temperature-dependent electrical resistivity ($\rho$) for (LaNdPrSmEu)$_{1-x}$Sr$_x$MnO$_3$ (0$\leq$\textit{x}$\leq$0.5) is shown in Fig.~5. (LaNdPrSmEu)$_{1-x}$Sr$_x$MnO$_3$ with \textit{x}=0.0 and 0.1 show significantly high $\rho$ at temperatures below 150\,K, which decreases exponentially with an increase in the temperature, indicating an insulating nature across the temperature range studied. High-entropy (LaNdPrSmEu)$_{1-x}$Sr$_x$MnO$_3$ samples exhibit insulating behavior for \textit{x} $\leq$ 0.2. It is noted that $\rho$ of (LaNdPrSmEu)MnO$_3$ is higher than that of LaMnO$_3$. This rise in $\rho$ for pristine sample may be understood using the one electron bandwidth $W$ of e$_g$ electrons which depends on the Mn-O-Mn bond angle as follows: $W \propto \frac{Cos \frac{1}{2}(\pi-<\theta>)}{d^{3.5}}$ where $<\theta>$ is the average Mn-O-Mn bond angle, and $d$ is the average Mn-O bond length.\cite{R49} Since the rare-earth site in the high-entropy sample is occupied with five different rare-earth elements having an average ionic radius smaller than La, it results in a decrease in the average ionic radii of the A-site and hence results in reduced Mn-O-Mn bond angle. This reduction further results in narrower one-electron bandwidth, favoring an insulating state. $\rho$ decreases with the Sr-substitution further and a similar nature of temperature-dependence of $\rho$ is obtained for the Sr-substitution at the rare-earth site up to \textit{x}=0.2. This decrease in electrical resistivity is attributed to the hopping of electrons from Mn$^{3+}$ to Mn$^{4+}$ via O$^{2-}$. The magnetic measurements for (LaNdPrSmEu)$_{1-x}$Sr$_x$MnO$_3$ with \textit{x}=0.2 show dominant FM ordering, whereas $\rho$-T indicates an insulating behavior. With increased Sr substitution ((LaNdPrSmEu)$_{0.7}$Sr$_{0.3}$MnO$_3$), $\rho$ decreases further and shows a metallic nature at low temperatures with metal-to-insulator transition (MIT) around $\sim$ 180\,K, which increases to 230\,K for (LaNdPrSmEu)$_{1-x}$Sr$_x$MnO$_3$ with \textit{x}=0.4. It is noted that the MIT transition temperature is close to the Curie temperature (T$_C$) obtained in these samples and hence shows the homogenous nature of the sample.\cite{R43} MIT is sharply seen in the sample with \textit{x}=0.3, however, the transition temperature is smaller than the one observed in La$_{0.7}$Sr$_{0.3}$MnO$_3$, which could be ascribed to the decreased ionic radii. The MIT vanishes for \textit{x}=0.5, which could be attributed to the chemical inhomogeneity suggested by the XRD analysis, or marginal charge-ordered state appearing at \textit{x}=0.5 for R$_{1-x}$Sr$_x$MnO$_3$, or both. This indicates that the Sr substitution at the rare-earth site in the high-entropy sample changes the resistive nature from insulating (\textit{x}=0.0) to metallic (\textit{x}=0.3) and is back to insulating for \textit{x}=0.5. In simple rare-earth manganites, La$_{1-x}$Sr$_x$MnO$_3$ shows insulator to metal crossover at \textit{x}=0.18\cite{R50}, Nd$_{1-x}$Sr$_x$MnO$_3$ changes from an insulator to metal for \textit{x}=0.28 and again becomes insulating at \textit{x}=0.44 due to charge-ordering, and it remains insulating for Pr-based manganites.\cite{R51} The changes in electrical resistivity as a function of Sr concentration are shown in the inset of Fig.~5. The electrical resistivity decreases with Sr substitution up to \textit{x}=0.4 and increases for \textit{x}=0.5 at 300\,K.\\
The electrical resistivity ($\rho$) and magnetoresistance (MR) of polycrystalline (LaNdPrSmEu)$_{1-x}$Sr$_x$MnO$_3$ (0 $\leq$ \textit{x} $\leq$ 0.5) pellets have been measured in the 2–300\,K range in magnetic fields up to 6\,T, shown in Fig.~6. Electrical resistivity remains unchanged under 6\,T magnetic field for the samples with \textit{x}=0.0 and 0.1 (Fig.~6(a)), indicating that the double exchange mechanism is not dominant in these samples. A notable decrease in electrical resistivity observed for \textit{x}=0.2 at 6\,T shows correlated magnetic-transport properties in the sample. The magnetization measurements depict an FM nature for \textit{x}=0.2 and hence suggest a ferromagnetic-insulator (FM-I) nature of the sample, a consequence of the competing DE and SE interactions. When increasing the Sr-content further, the application of magnetic field results in a significant decrease of the resistivity for (LaNdPrSmEu)$_{1-x}$Sr$_x$MnO$_3$ with \textit{x}=0.3 (Fig.~6(b)) and \textit{x}=0.4 (Fig.~6(c)), suggesting the ferromagnetic-metallic (FM-M) nature of the sample. It is seen that the magnetic field reduces the resistivity close to T$_C$ via shifting the resistivity maximum to a higher temperature. This originates from the suppression of the spin-scattering of mobile e$_g$ electrons by aligning the localized magnetic moments by an application of the magnetic field.\\
The magnetoresistance (MR) as a function of temperature for (LaNdPrSmEu)$_{1-x}$Sr$_x$MnO$_3$ (0$\leq$\textit{x}$\leq$0.5) is calculated using the experimentally measured values of resistivity under zero field ($\rho$(0)) and applied magnetic field ($\rho$(H)) using the following expression: $MR(\%)=[(\rho(H)-\rho(0))/(\rho(0))]\times100$ and is shown in Fig.~6(e). The magnitude of decrease in resistivity under the magnetic field is larger in regions close to T$_C$, or T$_p$ (the temperature at which the resistivity shows maxima). The resistivity decreases under 6\,T for \textit{x}=0.5 (Fig.~6(d)) at lower temperatures, and the change in MR\% decreases with increasing temperature and is attributed to possibly several competing magnetic contributions linked to the chemical inhomogeneity of the sample. An increase in MR is seen in the samples with increasing Sr concentration, correlated with the emergence of FM interactions, which is consistent with the studies related to simple rare-earth manganites.\cite{R52} A maximum MR\% of 90\% is obtained for \textit{x}=0.3 close to T$_C$. The variation of MR\% as a function of the magnetic field for \textit{x}=0.3-0.5 is shown in Fig.~6(f). A sharp decrease in MR\% is observed at low fields ($<$1T), followed by a more gradual change.\\
\\
\section{Conclusion}
The structural, magnetic, and electrical properties of Sr-substituted high-entropy rare-earth manganite have been discussed. The Rietveld refinement of the X-ray diffraction data suggests a structural transition from the orthorhombic (\textit{Pbnm}) phase to rhombohedral (\textit{R-3c}) phase with Sr-substitution at the rare-earth site for (LaNdPrSmEu)$_{1-x}$Sr$_x$MnO$_3$. Sr substitution at the rare-earth site enhances the Mn$^{4+}$ content in the sample, as confirmed by the X-ray photoelectron spectroscopy (XPS) analysis. (LaNdPrSmEu)$_{1-x}$Sr$_x$MnO$_3$ with \textit{x}=0.0 depicts a weak ferromagnetic insulator behavior. By increasing the Sr doping, namely hole-doping, a ferromagnetic-metallic state emerges, which is attributed to the dominant double exchange mechanism mediated by mobile carriers. The magnetic transition temperature for the high-entropy samples is lower than that of La$_{1-x}$Sr$_x$MnO$_3$, which can be explained by weaker magnetic interaction due to reduced ionic radii that result in buckling of Mn-O-Mn bonds. A correlation between the structural transition from orthorhombic to rhombohedral phase to magnetic and transport properties is observed. A strong dependence of metal-to-insulator transition and the Curie temperature on the average ionic radii of the A-site is observed. The magnitude of the negative magnetoresistance is larger close to the Curie temperatures, which is attributed to the suppression of spin scattering of e$_g$ carriers. This study demonstrates the strong correlation between magneto-transport properties in hole-doped high-entropy rare-earth manganites and paves directions to tune multifunctional properties in high-entropy materials via doping.\\

\section{Acknowledgments}
This work was supported by the French Agence Nationale de la Recherche (ANR), through the project NEO (ANR 19-CE30-0030-01).\\

\textbf{Note}: During the redaction of this manuscript, an article that discusses the magnetic and electronic properties of related hole-doped (GdLaNdSm)$_{1-x}$Sr$_x$MnO$_3$ (\textit{x}=0 - 0.5) oxides, has been accepted for publication. \cite{R53}\\
\end{document}